\documentclass[twocolumn,aps,superscriptaddress,amsmath,amssymb,preprintnumbers,floatfix]{revtex4-1}

\usepackage[utf8]{inputenc}
\usepackage{newtxtext}
\usepackage[upint]{newtxmath}
\usepackage{microtype}
\usepackage{textcomp}
\usepackage{eucal}
\usepackage{bm}

\usepackage{enumerate}
\usepackage{amsfonts}
\usepackage{amsmath}
\usepackage{amssymb}
\usepackage{color}
\usepackage{soul}

\usepackage{graphicx}

\usepackage{amsmath,amssymb}
\usepackage{graphicx,xcolor,soul,sidecap}
\usepackage[colorlinks,bookmarks=false,citecolor=blue,linkcolor=blue,urlcolor=blue,hyperfootnotes=true]{hyperref}
\usepackage[capitalize]{cleveref}

\makeatother

\begin{document}
\title{Controllable enhancement of $p$-wave superconductivity via magnetic coupling \\ to a conventional superconductor}

\author{Linde A. B. Olde Olthof} 
\email[]{labo2@cam.ac.uk}
\affiliation{Department of Materials Science \& Metallurgy, University of Cambridge, CB3 0FS Cambridge, United Kingdom}
\author{Lina G. Johnsen} 
\affiliation{Center for Quantum Spintronics, Department of Physics, Norwegian University of Science and Technology, NO-7491 Trondheim, Norway}
\author{Jason W. A. Robinson} 
\affiliation{Department of Materials Science \& Metallurgy, University of Cambridge, CB3 0FS Cambridge, United Kingdom}
\author{Jacob Linder} 
\email[]{jacob.linder@ntnu.no}
\affiliation{Center for Quantum Spintronics, Department of Physics, Norwegian University of Science and Technology, NO-7491 Trondheim, Norway}
\date{\today}

\begin{abstract}
Unconventional superconductors are of high interest due to their rich physics, a topical example being topological edge-states associated with $p$-wave superconductivity. A practical obstacle in studying such systems is the very low critical temperature $T_\text{c}$ that is required to realize a $p$-wave superconducting phase in a material. We predict that the $T_\text{c}$ of an intrinsic $p$-wave superconductor can be significantly enhanced by coupling it via an atomically thin ferromagnetic layer (F) to a conventional $s$-wave or a $d$-wave superconductor with a higher critical temperature. We show that this $T_\text{c}$-boost is tunable via the direction of the magnetization in F. Moreover, we show that the enhancement in $T_\text{c}$ can also be achieved using the Zeeman-effect of an external magnetic field. Our findings provide a way to increase $T_\text{c}$ in $p$-wave superconductors in a controllable way and make the exotic physics associated with such materials more easily accessible experimentally. 
\end{abstract}
\maketitle

\textit{Introduction}. Superconductivity is one of the most exotic states of matter and is often described as conventional or unconventional, depending on the symmetry of the underlying order parameter. Conventional spin-singlet superconductors have a $s$-wave order parameter that is isotropic in momentum space. Unconventional superconductors can instead have a highly anisotropic order parameter, both in magnitude and in phase. The spin-triplet $p$-wave ($p_x+ip_y$) order parameter is a prototypical example \cite{maeno_rmp_03}, which is of high interest due to its edge-states. Such edge-states may arise at interfaces of unconventional superconductors where reflection causes the order parameter to change sign \cite{Hu,Nagai1986,Nagai1995} with energies lying midgap at the normal-state Fermi level. Previous work has shown that edge-states arising from a $p$-wave superconductor can be topologically protected from decoherence \cite{Kitaev,Green,Gnezdilov}, making them interesting as building blocks for qubits in topological quantum computation \cite{Beenakker,Sato}.

Candidate materials for topological superconductivity include $^3$He B-phase \cite{Zhang}, the surface of Sr$_2$RuO$_4$ \cite{Tada}, Cu-doped Bi$_2$Se$_3$ \cite{Berg,Fu,Ando}, $p$-type TlBiTe$_2$ \cite{Yan} and BC$_3$ \cite{Jun}. Sr$_2$RuO$_4$ is the most studied although the exact underlying superconducting order parameter remains hotly debated \cite{Maeno2003,Ishida,Kallin}.
Sr$_2$RuO$_4$ has a critical temperature $T_\text{c}$ of 1.5~K \cite{Lichtenberg} and is known as one of the most disorder-sensitive superconductors \cite{Maeno1998}, making it challenging to utilize.

A way to locally increase the $T_\text{c}$ of Sr$_2$RuO$_4$ is via the 3~K phase, which involves embedding Ru inclusions into Sr$_2$RuO$_4$ \cite{Nakatsuji}. Similar local $T_\text{c}$ enhancement has been predicted near dislocations \cite{Liu2013}. 
The 3~K phase was later attributed to local stress induced by the Ru inclusions \cite{Schilfgaarde} and are mimicked in pure Sr$_2$RuO$_4$ by applying uniaxial pressure \cite{Maeno2009,Maeno2010}.
Piezoelectric-based techniques achieve an even higher compression and raise the $T_\text{c}$ globally to 3.4~K \cite{MacKenzie,Hicks}. 
By linking the uniaxial strain to spin and charge fluctuations, the latter could serve as a further mechanism for increasing $T_\text{c}$ \cite{Schilfgaarde}.

Finding a general method to enhance $T_\text{c}$ of unconventional superconductors in order to more easily access their interesting physics is an important, yet challenging goal. Proximity enhancement of $T_\text{c}$ has been demonstrated in $s$-wave systems \cite{Bishop,Gray,Chi,Parmenter,Aronov,Blamire}, finding applications in electronic cooling \cite{Klapwijk,Shurakov}.
However, in a junction between a singlet and triplet superconductor, there is no enhancement of the critical temperature in the low-$T_\text{c}$ superconductor, since singlet Cooper pairs do not couple to triplet pairs and vice versa \cite{Fenton1980,Millis}. 

In the case of transition-metal compounds, strong intra-ionic spin-orbit coupling can enhance triplet superconductivity in a three-layer oxide heterostructure \cite{horsdal_prb_16}. Hence, to couple singlet and triplet superconductors, a spin-active interface is required to facilitate conversion between singlet and triplet Cooper pairs. Ferromagnets \cite{Bergeret2001,Buzdin,LinderRobinson} and spin-orbit coupling \cite{Rashba,Annunziata,Bergeret2013} are commonly used for generating spin-triplet Cooper pairs from conventional superconducting pairing. In particular, both ferromagnets \cite{Tanaka,Yokoyama,Manske} and spin-orbit coupling \cite{Hasegawa,Kashiwaya} have been used to study the Josephson effect in $s$-wave/$p$-wave (S/P) junctions.

In this {\it Letter}, we present a method to boost $T_\text{c}$ of a triplet superconductor. The key is to couple a low-$T_\text{c}$ triplet superconductor to a higher-$T_\text{c}$ spin-singlet superconductor (either $s$-wave or $d$-wave) via a ferromagnetic interface (F). This is achievable using different types of ferromagnets, but here we consider an atomically thin ferromagnetic interlayer. Using numerical diagonalization of a lattice-model, we predict that such a coupling strongly enhances the $T_\text{c}$ of the triplet superconductor. Moreover, we show that the $T_\text{c}$-boost is controllable by rotating the ferromagnetic exchange field. Finally, we show that the enhancement of $T_\text{c}$ is also obtained via a Zeeman effect from an external magnetic field.

\begin{figure}
    \centering
    \includegraphics[width=\linewidth]{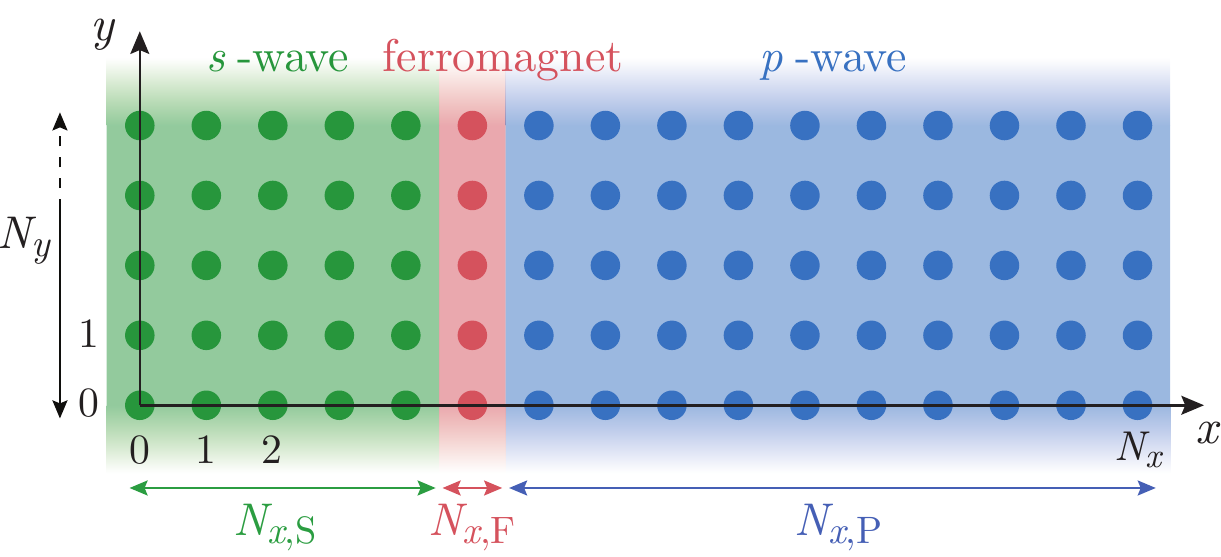}
    \caption{Schematic illustration of the two-dimensional S/F/P cubic $N_x\times N_y$ lattice structure, with layer thicknesses of $N_{x,\text{S}}$, $N_{x,\text{F}}$ and $N_{x,\text{P}}$ lattice sites, respectively. The $y$-direction is translationally invariant by using periodic boundary conditions and $N_y \gg N_x$.}
    \label{Fig1}
\end{figure}

\textit{Model}. We model a two-dimensional S/F/P junction using the tight-binding Bogoliubov-de Gennes framework \cite{Zhu,Kuboki2004,Terrade,JohnsenPRB} with a square $N_x \times N_y$ lattice structure \footnote{We expect the results to be qualitatively the same for a rectangular lattice or a three-dimensional model.}, as illustrated in Fig.~\ref{Fig1}. The interface normal is along the $x$-axis and we assume periodic boundary conditions along $y$.
The Hamiltonian in terms of the second quantization electron creation and annihilation operators $c_{\boldsymbol{i}\sigma}^\dagger$ and $c_{\boldsymbol{i}\sigma}$ is
\begin{align}
    H &= - t \sum_{\langle \boldsymbol{i,j}\rangle, \sigma} c_{\boldsymbol{i}\sigma}^\dagger c_{\boldsymbol{j}\sigma} 
    - \sum_{\boldsymbol{i},\sigma} \mu_{\boldsymbol{i}} n_{\boldsymbol{i}\sigma} 
    - \sum_{\boldsymbol{i}} U_{\boldsymbol{i}} n_{\boldsymbol{i}\uparrow} n_{\boldsymbol{i} \downarrow} \nonumber \\
    & - \frac{1}{2} \sum_{\langle \boldsymbol{i,j} \rangle ,\sigma} V_{\boldsymbol{ij}} n_{\boldsymbol{i}\sigma}n_{\boldsymbol{j},-\sigma} 
    + \sum_{\boldsymbol{i},\sigma,\sigma'} c_{\boldsymbol{i}\sigma}^\dagger (\boldsymbol{h}_{\boldsymbol{i}}\cdot\boldsymbol{\sigma})_{\sigma\sigma'} c_{\boldsymbol{i}\sigma},
\end{align}
where $\boldsymbol{i}=(i_x,i_y)$ is the lattice site, $t$ is the hopping amplitude, $\mu_{\boldsymbol{i}}$ is the chemical potential and $n_{\boldsymbol{i}\sigma} \equiv c_{\boldsymbol{i}\sigma}^\dagger c_{\boldsymbol{i}\sigma}$ is the number operator. The attractive on-site interaction ($U_{\boldsymbol{i}} >0$) gives rise to isotropic singlet superconductivity in S, while the nearest-neighbour interaction ($V_{\boldsymbol{ij}} >0$) between opposite spin electrons at sites $\boldsymbol{i}$ and $\boldsymbol{j}$ results in spin-triplet superconductivity in P.
The last term describes a spin-splitting field $\boldsymbol{h}_i$ interacting with the Pauli spin matrices $\boldsymbol{\sigma}=(\sigma_x,\sigma_y,\sigma_z)$ to give a net spin polarization of the itinerant electrons in F.

The two superconducting terms are treated by a mean-field approach, assuming $c_{\boldsymbol{i}\uparrow}c_{\boldsymbol{i}\downarrow} = \langle c_{\boldsymbol{i}\uparrow}c_{\boldsymbol{i}\downarrow}\rangle +\delta$ and neglecting second order fluctuations in $\delta$. 
We obtain one on-site pair correlation $F_{s,\boldsymbol{i}} = \langle c_{\boldsymbol{i}\uparrow} c_{\boldsymbol{i}\downarrow} \rangle$ and four nearest-neighbour pair correlations $F_{\boldsymbol{i,i}\pm\hat{x}} = \langle c_{\boldsymbol{i}\uparrow} c_{\boldsymbol{i}\pm \hat{x},\downarrow} \rangle$ and $F_{\boldsymbol{i,i}\pm\hat{y}}= \langle c_{\boldsymbol{i}\uparrow} c_{\boldsymbol{i}\pm \hat{y},\downarrow} \rangle$. These are anomalous Green functions quantifying the strength of the superconducting correlations in the material and vanish at $T_\text{c}$. $\hat{x}$ and $\hat{y}$ are vectors connecting nearest-neighbor sites along the $x$ and $y$ axis, respectively.
The Hamiltonian is diagonalized numerically and solved iteratively for these five pair correlations. 
To describe the triplet symmetry, we introduce the symmetrized nearest-neighbour triplet pair correlation $F_{\boldsymbol{ij}}^\text{(T)} = (F_{\boldsymbol{ij}}-F_{\boldsymbol{ji}})/2$. After convergence, we calculate the superconducting order parameters
\begin{align}
    & \Delta_{s,\boldsymbol{i}} = U F_{s,\boldsymbol{i}}, \\
    & \Delta_{p_x,\boldsymbol{i}} = V F_{p_x,\boldsymbol{i}} = \frac{V}{2}\big( F_{\boldsymbol{i},\boldsymbol{i}+\hat{x}}^{\text{(T)}} - F_{\boldsymbol{i},\boldsymbol{i}-\hat{x}}^{\text{(T)}}  \big), \\
    & \Delta_{p_y,\boldsymbol{i}} = V F_{p_y,\boldsymbol{i}} = \frac{V}{2}\big( F_{\boldsymbol{i},\boldsymbol{i}+\hat{y}}^{\text{(T)}} - F_{\boldsymbol{i},\boldsymbol{i}-\hat{y}}^{\text{(T)}} \big).
\end{align}
The anomalous Green functions $F_{s,\boldsymbol{i}}$, $F_{p_x,\boldsymbol{i}}$ and $F_{p_y,\boldsymbol{i}}$ have their own critical temperatures $T_\text{c}^s$, $T_\text{c}^{p_x}$ and $T_\text{c}^{p_y}$, respectively, in the sense that they become smaller than some tolerance level at a specific temperature. The highest $T_\text{c}$ of an anomalous Green function determines the temperature at which the material becomes superconducting.
The full derivation of the model is given in~\cite{supp}.
The nearest-neighbour model can describe a variety of superconducting symmetries. To stabilize the $p$-wave pairing, the $V$ and $\mu$ (inside P) parameters are chosen in accordance with the free energy minimization in Ref.~\cite{Kuboki2001}. In the following, we study $F_{p_x}$ and $F_{p_y}$ to determine how the coupling between the superconductors in Fig. \ref{Fig1} through an atomically thin ferromagnet influences the triplet superconductivity.

\begin{figure}
    \centering
    \includegraphics[width=\linewidth]{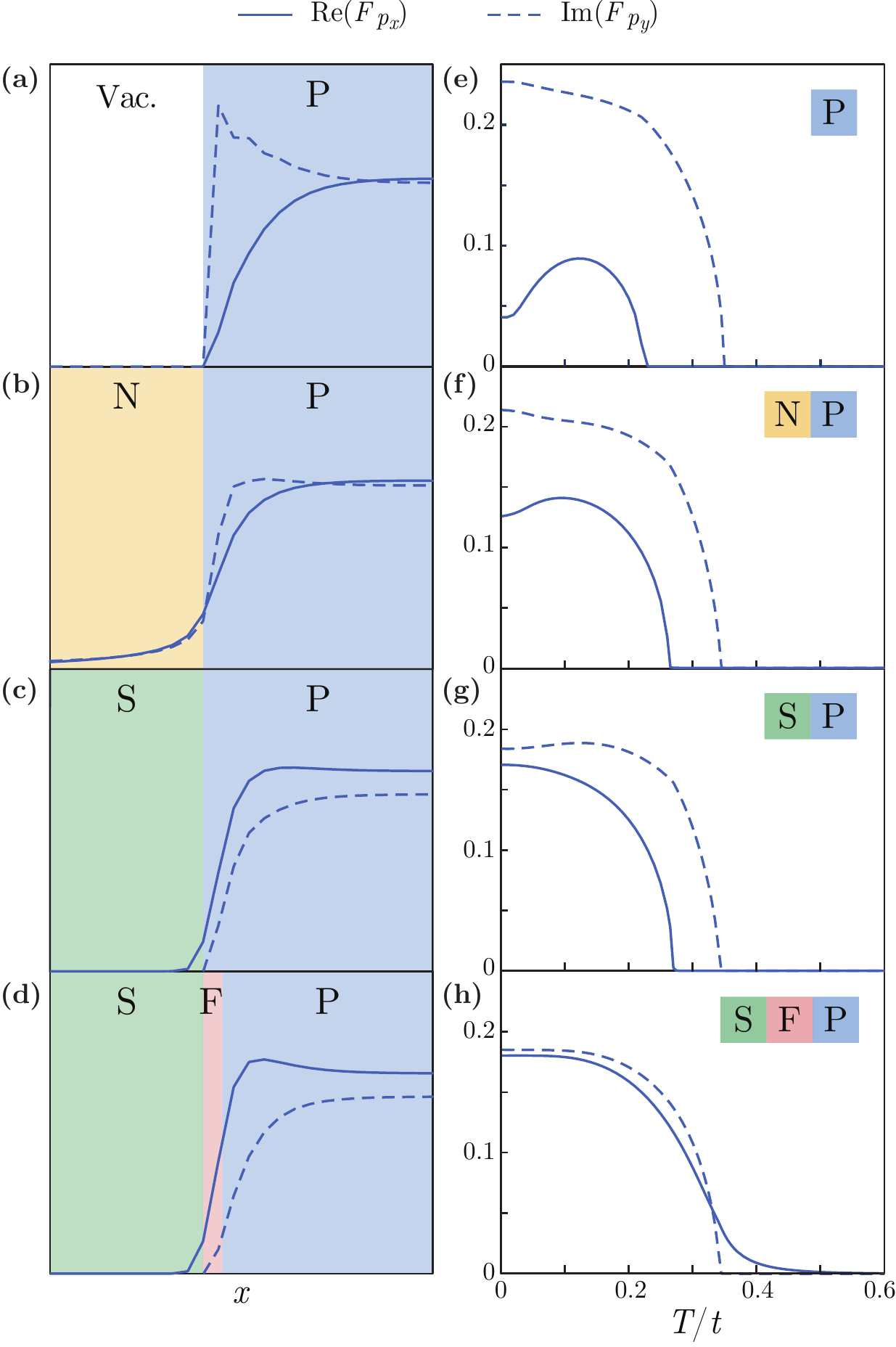}
    \caption{Spatial pair correlation profiles Re($F_{p_x}$) (solid) and Im($F_{p_y}$) (dashed) at the interface of {\bf (a)} vacuum/P, {\bf (b)} N/P,  {\bf (c)} S/P and {\bf (d)} S/F/P at zero temperature. The pair correlations vs normalized temperature for {\bf (e)} thin single P, {\bf (f)} thin-layer N/P, {\bf (g)} S/P and {\bf (h)} S/F/P junctions ($N_{x,\text{N}} = N_{x,\text{S}} = 5$, $N_{x,\text{F}} = 1$, $N_{x,\text{P}} = 10$). The thin single P is severely suppressed. The suppression is recovered in N/P and S/P. Singlet-triplet conversion in S/F/P results in a tail in $T_\text{c}$. The parameters for all graphs are $N_y=200$, $\mu_\text{N}/t = \mu_\text{S}/t = 1.2$, $U/t = 5.3$, $\mu_\text{F}/t = 1.4$, $h_z/t = 0.9$, $\mu_\text{P}/t = 1.8$ and $V/t = 1.5$.}
    \label{Fig2}
\end{figure}

\textit{$T_\text{c}$-boost via singlet-triplet coupling}. We first consider the pair correlations close to the surface of a finite two-dimensional $p_x+ip_y$ superconductor shown in Fig. \ref{Fig2}(a).
In a single P (interfaced with vacuum), electrons are reflected with opposite momentum in the $x$-direction ($k_x \mapsto -k_x$), while the momentum in the $y$-direction is conserved.
The $p_x$ orbital symmetry $\sin(k_x)$ is odd under inversion of $k_x$. This symmetry relation $\Delta(k_x ) = -\Delta(-k_x )$ gives the criterion for having a midgap surface state \cite{Hu,Nagai1995}. A Cooper pair thus experiences an opposite sign after reflection. This causes the pair breaking that gives the cancellation of $F_{p_x}$ near the surface \cite{Rainer,Nagai1986}. As a result, there are more electrons present to form Cooper pairs in the $y$-direction and $F_{p_y}$ increases close to the surface.

By bringing the P into contact with a normal metal (N) in Fig. \ref{Fig2}(b), 
$F_{p_x}$ is no longer fully cancelled at the interface since Cooper pairs can enter the N via the proximity effect, thus inhibiting the formation of midgap states at the surface, since not all electrons are reflected any longer. On the other hand $F_{p_y}$ simply decreases at the interface since Cooper pairs can now tunnel into N. Both $F_{p_x}$ and $F_{p_y}$ decay exponentially in N. 
In Fig.~\ref{Fig2}(c), we replace N with a conventional superconductor S, forming a S/P junction. The proximity effect is strongly suppressed \cite{Fenton1980,Millis} and spans only a few lattice sites on either side of the interface. Consequently, $F_{p_x}$ and $F_{p_y}$ reach their bulk values close to the interface.
Like the N/P case, midgap surface state reflections are also suppressed in S/P \footnote{Replacing the vacuum with any material suppresses the midgap surface state since the reflection probability goes from 1 to $<1$.} and $F_{p_x}$ overtakes $F_{p_y}$. 
Finally, by sandwiching a F in between the S and P, as shown in Fig. \ref{Fig2}(d), conversion of $s$-wave singlet into $p_x$-wave triplets takes place and $F_{p_x}$ is boosted at the interface. Increasing the exchange field results in an enhancement of singlet-to-triplet pair conversion efficiency. Additionally, increasing the field weakens the influence of the midgap surface state \cite{Tanaka}, thus strengthening $F_{p_x}$. 

Having demonstrated the behavior of the triplet correlations $F_{p_x}$ and $F_{p_y}$ in different types of heterostructures in Figs. \ref{Fig2}(a)-(d), we now consider the temperature dependence of these correlations in order to demonstrate that $T_\text{c}$ of the triplet superconductor can be enhanced. First, consider a system where P is thin ($N_{x,\text{P}}$ small) enough that the midgap surface states of the two surfaces partially overlap.
If P is interfaced by vacuum on either side, $F_{p_x}$ vanishes at both interfaces and is severely or even fully suppressed over the whole width of the superconductor. Hence, its $T_\text{c}$ (taken in the middle of the P) is suppressed as well, as shown in the purple graph in Fig.~\ref{Fig2}(e).
The behavior of $F_{p_x}(T)$ is non-monotonic which is a known result in the presence of midgap surface states in a thin P \cite{Nagai1986,Nagai1995,Vorontsov}. Placing the P in contact with an N or S instead of a vacuum, $F_{p_x}$ can be recovered by reducing midgap surface state reflections, which is visible in Fig.~\ref{Fig2}(f)-(g) for the S/P junction. The $T_\text{c}$ in this case matches the $T_\text{c}$ of a bulk P.

The S/F/P junction shows an additional effect. The S has a higher $T_\text{c}$ than P; for our parameters, $T_\text{c}^s \approx 10 T_\text{c}^{p_x}$. Once the intrinsic $T_\text{c}$ of P is exceeded, there is still a small amount of triplets coming from the S/F interface, stabilising $F_{p_x}$ above its intrinsic $T_\text{c}$. This results in a tail in $T_\text{c}^{p_x}$, as seen in Fig.~\ref{Fig2}(h). For the optimal parameters, it is possible to nearly double $T_\text{c}^{p_x}$. Only $T_\text{c}^{p_x}$ is boosted while $T_\text{c}^{p_y}$ remains the same as a result of the structural symmetry. 
In this setup, the interface normal is parallel to the $x$ axis, meaning spatial inversion symmetry is broken along $x$ and F only converts $s$-wave singlets into $p_x$-wave triplets \footnote{To be precise: the S/F bilayer converts even-frequency $s$-wave singlets into even-frequency $p_x$-wave triplets and odd-frequency $s$-wave triplets \cite{JohnsenPRL}. The latter do not contribute to the $T_\text{c}$-enhancement discussed here}.
Since there is no symmetry breaking in the $y$-direction, there is no conversion to $p_y$-wave triplets \cite{JohnsenPRL}.
Furthermore, the increase in the magnitude of $F_{p_x}$ is accompanied with a decrease in $F_{p_y}$ since less midgap surface states appear at the P interface.

The increase in $T_\text{c}^{p_x}$ is limited by the $T_\text{c}$ of the superconductor with the highest $T_\text{c}$ in the heterostructure and by the interface transparency. The effective interface transparency depends on the presence of an explicit barrier (not included in our model) and the Fermi surface mismatch, i.e. different choices of $\mu$. 
There is no intrinsic enhancement of the pairing mechanism, since $U$ and $V$ stay constant throughout our calculations.
Regarding the layer thicknesses, $T_\text{c}^{p_x}$ is doubled in S/F/P compared to a thin single P with suppressed $T_\text{c}^{p_x}$. The same effect is expected in thicker layers, although the absolute $T_\text{c}^{p_x}$ increase will be smaller.

It is known that triplet superconductivity is also induced in S/F bilayers \cite{Bergeret2001,Yokoyama}. However, both singlet and triplet correlations occur simultaneously in these systems, leading to mixed-pairing superconductivity. In contrast, in our case only the triplet correlations have a non-negligible magnitude in the temperature-regime exceeding the intrinsic $T_\text{c}^{p_x}$, distinguishing it from the S/F bilayer.

\begin{figure}
    \centering
    \includegraphics[width=\linewidth]{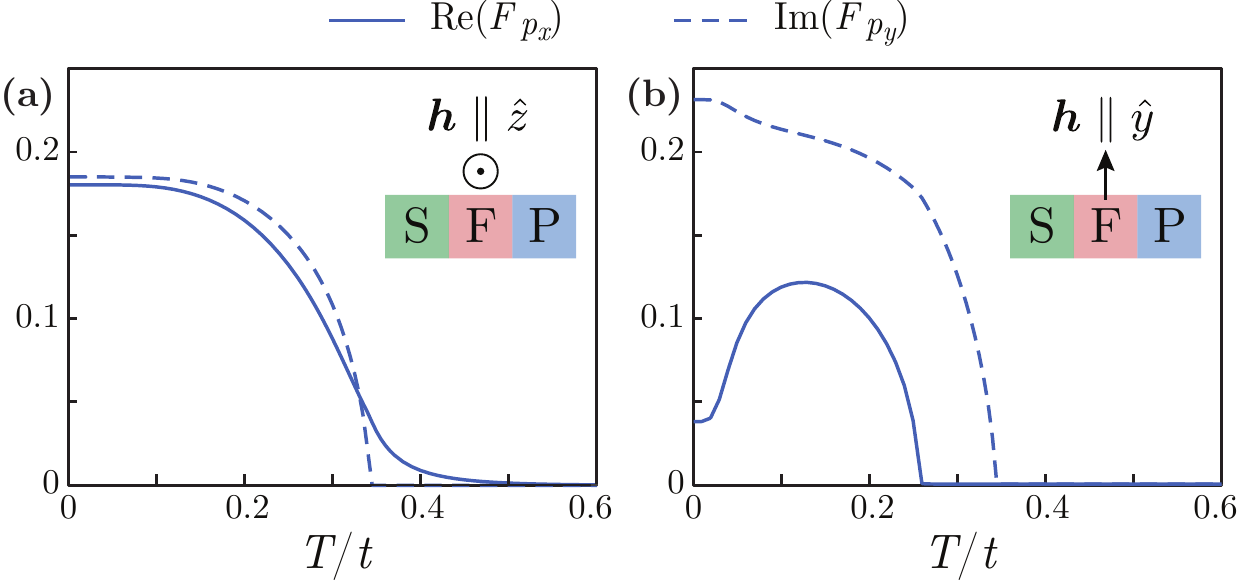}
    \caption{The pair correlations vs normalized temperature for S/F/P with the exchange field in F {\bf (a)} along $\hat{z}$ and {\bf (b)} along $\hat{y}$. Rotating the exchange field changes $T_\text{c}$ dramatically. The parameters are $N_{x,\text{S}} = 5$, $N_{x,\text{F}} = 1$, $N_{x,\text{P}} = 10$, $N_y=200$, $\mu_\text{S}/t = 1.2$, $U/t = 5.3$, $\mu_\text{F}/t = 1.4$, $h_z/t = h_y/t = 0.9$, $\mu_\text{P}/t = 1.8$ and $V /t= 1.5$.}
    \label{Fig3}
\end{figure}

\textit{Controlling $T_\text{c}$-enhancement via magnetization direction.} 
Triplet superconductivity is generally described by the $\boldsymbol{d}$-vector $\boldsymbol{d} \equiv [ (\Delta_{\downarrow\downarrow} - \Delta_{\uparrow\uparrow})/2, -i(\Delta_{\uparrow\uparrow} + \Delta_{\downarrow\downarrow})/2, \Delta_{\uparrow\downarrow}$] \cite{LinderRobinson}. 
We consider $p_x+ip_y$ pairing happens between opposite spin electrons so that $\boldsymbol{d}$ is along $\hat{z}$.
We study the effect of changing the direction of the F spin-splitting field $\boldsymbol{h}$ with respect to $\boldsymbol{d}$.
The S is isotropic and the $\mid \uparrow\downarrow \rangle - \mid \downarrow \uparrow \rangle$ spin-singlet Cooper pair is rotationally invariant. 
The three spin-triplet states $\mid \uparrow\downarrow \rangle + \mid \downarrow \uparrow \rangle$, $\mid \uparrow\uparrow \rangle$ and $\mid \downarrow \downarrow \rangle$ transform into each other when the quantization axis changes \cite{Eschrig,LinderRobinson}, and we choose this axis to be along $\boldsymbol{h}$.
Thus, a spin-splitting field $h_z$ converts singlets to $\mid \uparrow\downarrow \rangle + \mid \downarrow \uparrow \rangle$ triplets polarized along the $z$ axis. This is the native triplet Cooper pairs of the P and hence boosts $T_\text{c}^{p_x}$. By changing the exchange field direction from $h_z$ to $h_x$ or $h_y$, singlets are converted to $\mid \uparrow\downarrow \rangle + \mid \downarrow \uparrow \rangle$ triplets with a quantization axis along the $x$ and $y$ axis, respectively. In the reference frame of the $p_x + ip_y$-wave superconductor, this corresponds to $\mid \uparrow\uparrow \rangle$ and $\mid \downarrow \downarrow \rangle$ triplets. These triplets suppress $T_\text{c}^{p_x}$, as seen in Fig.~\ref{Fig3}. For the optimal parameters, $T_\text{c}^{p_x}$ corresponding to $h_z$ is double the $T_\text{c}^{p_x}$ for $h_x$ or $h_y$. Vice versa, the F also converts triplets originating from P into singlets. Since the S is isotropic, this contribution is independent of the exchange field direction. 

The fact that the $T_\text{c}$-enhancement is controlled by the magnetization direction of $F$ is an important observation which could lead to interesting device concepts and a means to further probe $p$-wave superconductivity. The exchange field direction is tunable via an applied magnetic field, meaning that $T_\text{c}^{p_x}$ can be tuned externally, while $T_\text{c}^s$ remains largely unchanged. This can serve as a switch in a device. Similarly, to observe a Josephson current in a S/F/P junction, an exchange field component parallel to the $\boldsymbol{d}$-vector of the triplet order parameter (here pointing along $z$) is required \cite{Manske}. By extension, one could control the Josephson current by rotating an applied magnetic field.

\begin{figure}
    \centering
    \includegraphics[width=\linewidth]{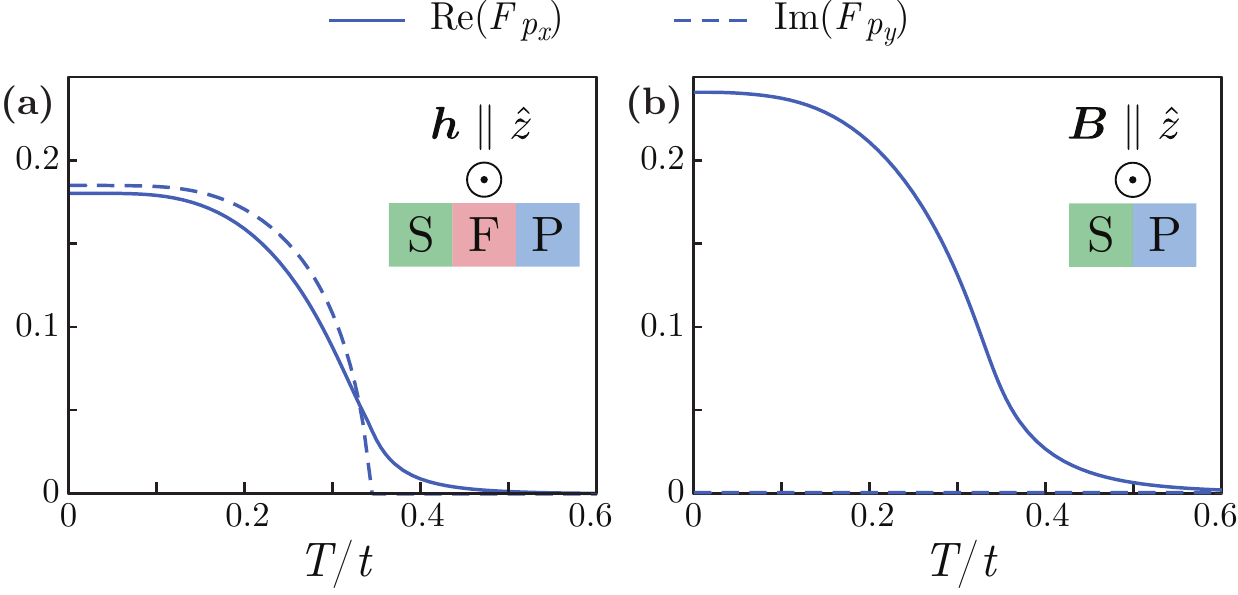}
    \caption{Comparison between the pair correlations in a S/F/P {\bf (a)} with exchange field $h_z$ and {\bf (b)} S/P in an external field $B_z$, vs normalized temperature. The external field is chosen as $B_z/t = B_\text{c}^{p_y} =  0.3$, for which $F_{p_x}$ is maximised.
     The parameters are $N_{x,\text{S}} = 5$, $N_{x,\text{F}} = 1$, $N_{x,\text{P}} = 10$, $N_y=200$, $\mu_\text{S}/t = 1.2$, $U/t = 5.3$, $\mu_\text{F}/t = 1.4$, $h_z/t = 0.9$, $\mu_\text{P}/t = 1.8$ and $V/t = 1.5$. }
    \label{Fig4}
\end{figure}

\textit{$T_\text{c}$-boost via external magnetic field.} Finally, we compare the S/F/P junction to a S/P junction with an external field $B_z$ along $\hat{z}$. When the superconductors are much smaller than the magnetic penetration depth $\lambda$, the orbital effect of $B_z$ is quenched and superconductivity coexists with a Zeeman-splitting throughout both superconductors up to the Clogston-Chandrasekhar limit \cite{clogston_prl_62, chandrasekhar_apl_62}. This results in a spin polarization across the whole junction. 
Since $F_{p_x}$ and $F_{p_y}$ have different magnitudes and different $T_\text{c}$, they also have different critical fields $B_\text{c}^{p_x}$ and $B_\text{c}^{p_y}$, respectively, with $B_\text{c}^{p_x} > B_\text{c}^{p_y}$.

By applying the field, first $F_{p_y}$ decreases. Since less Cooper pairs are converted from $F_{p_x}$ to $F_{p_y}$, this goes accompanied with a net increase in $F_{p_x}$.
Naturally, the $F_{p_x}$-to-$F_{p_y}$ conversion stops at $B_\text{c}^{p_y}$, at which $F_{p_x}$ reaches its maximum. This case is shown in Fig.~\ref{Fig4}.
Similar to $h_z$, $B_z$ facilitates singlet-to-triplet conversion and $T_\text{c}^{p_x}$ shows a tail.
Interestingly, the magnitude of $F_{p_x}$ in S/P with $B_z$ is significantly larger than in S/F/P, due to the lack of $F_{p_x}$-to-$F_{p_y}$ conversion.

However, $B_z$ also introduces a gradient of $F_s$ over the full width of the P (positive at the S/P interface, zero in the middle, negative at the P/vacuum interface). The magnitude of $F_s$ at the interfaces is approximately half the magnitude of $F_{p_x}$, which is significant, especially since the interesting physics unique to P superconductivity are generally situated at the edges. In this respect, the S/F/P structure is favorable since it maintains the pure $p$-wave correlations in the P throughout the regime of increased critical temperature. Increasing $B_z$ further, our model shows FFLO oscillations in the P.

Having demonstrated that $T_\text{c}$ is enhanced in a P by proximity to a S, it would be interesting to explore enhancing $T_\text{c}$ further using a high-$T_\text{c}$ cuprate $d_{x^2-y^2}$-wave superconductor (D). The theoretical framework used in this paper does not allow us to address the D case. The reason is that when solved self-consistently, as is required to compute $T_\text{c}$, our model tends to stabilise $d+p_x$ symmetry in D rather than pure $d$-wave  \cite{Kuboki2001}. However, our findings for the S/F/P case are indicative that a much larger $T_\text{c}^{p_x}$ could be possible if one is able to experimentally realize a high-$T_\text{c}$ superconductor/F/P junction.

Possible materials combinations with Sr$_2$RuO$_4$ include the transition metal ferromagnet SrRuO$_3$ \cite{SrRuO3}, the highly spin-polarized manganites such as La$_{0.7}$Ca$_{0.3}$MnO$_3$ \cite{LCMO,Anand} and the oxide superconductors YBa$_2$Cu$_3$O$_{7-x}$ \cite{YBCO} and Pr$_{1.85}$Ce$_{0.15}$CuO$_4$ \cite{PCCO}. Other materials to consider are two-dimensional ferromagnets including Cr$_2$Ge$_2$Te$_6$ \cite{CrGeTe}, CrI$_3$ \cite{CrI} and VSe$_2$ \cite{VSe}. 

\textit{Concluding remarks}. We have shown that the $T_\text{c}$ of a spin-triplet $p$-wave superconductor is controllable in a S/F/P junction, where S has a higher $T_\text{c}$ than P. A ferromagnetic interlayer facilitates singlet-to-triplet conversion, providing the P with triplets even above its intrinsic $T_\text{c}$. This shows up as a tail in the order parameter-temperature phase diagram.
Rotating the F exchange field direction with respect to the $p$-wave $\boldsymbol{d}$-vector controls the triplets, and therefore, $T_\text{c}$. An exchange field parallel to $\boldsymbol{d}$ is able to nearly double $T_\text{c}$, whereas an exchange field perpendicular to $\boldsymbol{d}$ converts singlets to the wrong type of triplets and suppresses $T_\text{c}$.
Hence, the exchange field direction serves as a $T_\text{c}$ switch and can by extension control a Josephson current.
In our model, we considered an atomically thin F. Qualitatively similar results are expected for thicker F.

Enhancing the $T_\text{c}$ of a $p$-wave superconductor above liquid helium temperatures would have massive practical advantages from a device operation point of view. In the case of Sr$_2$RuO$_4$ with its $T_\text{c}$ of 1.5~K, the doubling of $T_\text{c}$ is still not enough, although it might be sufficient for different $p$-wave materials. However, our results indicate that by replacing the $s$-wave with a high-$T_\text{c}$ cuprate $d$-wave superconductor the $p$-wave $T_\text{c}$ will be boosted even further.

\textit{Acknowledgements}. L.A.B.O.O. and J.W.A.R. were supported by the EPSRC through the Core-to-Core International Network ``Oxide Superspin''  (EP/P026311/1), the ``Superconducting Spintronics'' Programme Grant (EP/N017242/1), the Doctoral Training Partnership Grant (EP/N509620/1) and the Cambridge NanoDTC (EP/S022953/1). L.G.J. and J.L. were supported by the Research Council of Norway through its Centres of Excellence funding scheme grant 262633 QuSpin.

\bibliography{ref}

\end{document}